\documentclass{article}
\usepackage[utf8]{inputenc}
\usepackage[ruled,vlined]{algorithm2e}
\usepackage{amsmath}
\usepackage{amsthm}
\usepackage{amssymb}
\usepackage{graphicx}
\usepackage{array}
\usepackage{pgfplots}
\usepackage[tableposition=top]{caption}
\usepackage[a4paper, total={7in, 9in}]{geometry}
\usepackage{natbib}
\usepackage{placeins}

\pgfplotsset{width=10cm,compat=1.9}

\theoremstyle{definition}
\newtheorem{definition}{Definition}[section]

\newtheorem{theorem}{Theorem}[section]

\title{High Performance Level-1 and Level-2 BLAS}

\SetKw{KwBy}{by}
\newcommand{\assign}[2]{$#1 \gets #2$}

\author{Amit Singh \and Cem Bassoy}

\SetKw{KwBy}{by}

\begin{document}

\maketitle

\section{Introduction}

The introduction of the Basic Linear Algebra Subroutine (BLAS) in the 1970s laid the path to the different libraries to solve the same problem with an improved approach and improved hardware. The new BLAS implementation led to High-Performance Computing (HPC) innovation, and all the love went to the level-3 BLAS due to its humongous application in different fields not bound to computer science.However, level-1 and level-2 got neglected, and we here tried to solve the problem by introducing the new algorithm for the vector-vector dot product, vector-vector outer product and matrix-vector product, which improves the performance of these operations in a significant way.

We are not introducing any library but algorithms, which improves upon the current state-of-art algorithms. Also, we rely on the FMA instruction, OpenMP, and the compiler to optimize the code rather than implementing the algorithm in assembly. Therefore, our current implementation is machine oblivious and depends on the compiler's ability to optimize the code.
This paper makes the following contribution:

\begin{itemize}
    \item It compares the data movement and I/O lower bound of $GEMV$, $GER$, and $DOT$.
    \item It shows how to use multi-threads to utilize the modern multi-core hardware.
    \item It provides the shortcomings of our current implementation.
    \item It demonstrates the performance that is highly competitive with libraries such as MKL, BLIS, Eigen, and OpenBLAS.
\end{itemize}

\subsection{Notation}
Throughout the paper, we will use well-established symbols for describing the Linear Algebra objects. The uppercase Roman letters for the matrices (e.g. $A$, $B$, and $C$), the lowercase Roman letters for the vectors (e.g. $x$, $y$, and $z$), the Greek symbols for scalars (e.g. $\alpha$, $\beta$, and $\gamma$), and $i$,$j$, and $k$ for indices into the object.

\section{Machine Model}

The machine model that we will follow is similar to the model defined in the 
\citep{BLIS}, which takes modern hardware into mind. Such as vector registers 
and a memory hierarchy with multiple levels of set-associative data caches. However, 
we will add multiple cores functionality to the model and where each core has at least one cache level that not shared among the other cores.

All the data caches are set-associative and we can characterize them based 
on the four parameter defined bellow:

\(C_{L_i}\): cache line of the \(i^{th}\) level

\(W_{L_i}\): associative degree of the \(i^{th}\) level

\(N_{L_i}\): Number of sets in the \(i^{th}\) level

\(S_{L_i}\): size of the \(i^{th}\) level in Bytes

\begin{equation}
    S_{L_i} = C_{L_i}W_{L_i}N_{L_i}
    \label{eq:cache_size}
\end{equation}

Let the $S_{data}$ be the width of the type in bytes and $M_{L_i} = \frac{S_{L_i}}{S_{data}}$.

We are assuming that the cache replacement policy for all cache levels is 
\textbf{LRU}, which also assumed in the \citep{BLIS} and the cache 
line is same for all the cache levels. For most of the case, 
we will try to avoid the associative so that we could derive 
a simple equation containing the cache size only from the equation \ref{eq:cache_size}.
\section{I/O Lower Bound}
In this section, we will define the model of computation, which is precisely the same as defined in \cite{DBLP:journals/corr/SmithG17} for the problem $DOT$, $GEM$, and $GEMV$. We will use $A_{ij}$ for the element in $i^{th}$ row and $j{th}$ column, and $x_i$ for $i^{th}$ element. The $C$ or $c$ will denote the non-trivial partial sum matrix or vector.

\subsection{Problem Definition}
Here, we first begin defining the problem we have in hand for the operations with elementary instruction used in the algorithm. Our primary concern is multiplications, additions, and accumulation.

\begin{definition}[DOT]
The dot operation of $x$ and $y$ with the problem size $n$ computes $\alpha$ += $x \cdot y$ with elementary multiplications $\alpha_p$ = $x_p$ $y_p$, and we get $\alpha$ by summing over $\alpha_p$ for all $p$. In the end, we add initial $\alpha$. Therefore, there are $n$ elementary multiplications and additions in the operation.
\end{definition}

\begin{definition}[GEM]
The outer product of $x$ and $y$ with the problem size $m$ and $n$ computes $C$ += $x \otimes y$ with elementary multiplications $C_{ij}$ = $x_i$ $y_j$. In the end, we add initial $C_{ij}$. Therefore, there are $mn$ elementary multiplications and $0$ elementary additions in the operation.
\end{definition}

\begin{definition}[GEMV]
The matrix-vector product of $A$ and $x$ with the problem size $m$ and $n$ computes $c$ += $Ax$ with elementary multiplications $c_{p}$ = $A_{pj}$ $x_j$, and we get $c_i$ by summing over $c_p$ for all $p$. In the end, we add initial $c_i$. Therefore, there are $n$ elementary multiplications and $m-1$ elementary additions in the operation.
\end{definition}

\subsection{Model of Computation}
There are six operations defined in \cite{DBLP:journals/corr/SmithG17}, which we will also follow.

\begin{enumerate}
    \item \textbf{Read}. Move data from the slow memory into the fast memory.
    \item \textbf{Write}. Move data from the fast memory into the slow memory
    \item \textbf{Multiply}. If the input data is in the fast memory then multiply them and then create a variable in the fast memory to store the partial solution into it.
    \item \textbf{Add}. If the input data is in the fast memory then add them and create a variable in the fast memory to store the partial solution into it.
    \item \textbf{FMA}. If the input data is in the fast memory then multiply and add them without allocating additional space in the fast memory.
    \item \textbf{Delete}. Remove the variable from the fast memory if it is no longer needed.
\end{enumerate}

Before we get to the nitty-gritty details, we have to define $M$ and $R$.
$M$ is the total number of elements inside the fast memory, and $R$ is the total number of elements that need to bring inside the fast memory from the slow memory. $F_{M+R}$ is the total number of \textbf{FMA} instruction which will be issued.

To find the I/O lower bound of the operation, we will use the direct results from the paper \cite{DBLP:journals/corr/SmithG17}.

\begin{equation}
    TotalRead = (\frac{MemoryOperations}{F_{M+R}} - 1)R
    \label{eq:total_read}
\end{equation}

To minimize the total reads, we have to minimize the number of memory operations which is not possible because it is not in our but depends on the problem size. On the other hand, we can maximize the $F_{M+R}$, which is the only viable solution here.

From the paper \cite{INEQ}, we get the following inequality.

\begin{theorem}[Discrete Loomis-Whitney Inequality]
Let $V$ be a finite set with elements in $\mathbb{Z}^n$, and let $V_i$ be the orthogonal projection of $V$ onto the co-ordinate plane. Then the cardinality of $V$ is $|V|$, which satisfies
\[|V| = (|V_0||V_1| \hdots |V_n|)^{n-1}\]
\end{theorem}

Let $N_i$ be the number of elements inside the Linear Algebra object, then one can perform at least $|V|$ number of \textbf{FMA} operations.

\begin{equation}
    F_{M+R} \leq (|N_1| \hdots |N_n|)^{n-1}
\end{equation}

\begin{equation*}
    Maximize\ F_{M+R}\ under\ the\ constraint = 
    \begin{cases}
        F_{M+R} \leq (|N_0||N_1| \hdots |N_n|)^{n-1} \\
        N_1, N_2, \hdots, N_n \geq 0 \\
        N_1 + N_2 + \hdots + N_n \leq M + R
    \end{cases}
\end{equation*}

From the paper, in order to maximize the equation $N_1 + \hdots + N_n \leq M + R$, we use the Lagrange multiplier method, and we get
\[N_1 = \hdots = N_n = N\]
Therefore, $N = \frac{M + R}{n}$ and $F_{M+R} = (\frac{M + R}{n})^{\frac{n}{n-1}}$. Now, we back substitute $F_{M+R}$ into equation \ref{eq:total_read}.

\begin{equation*}
    TotalRead = (\frac{n^{\frac{n}{n-1}} MemoryOperations}{(M + R)^{\frac{n}{n-1}}} - 1)R
\end{equation*}

The maximum occurs when $R = 2M$.

\begin{equation*}
    TotalRead = 2M(\frac{n^{\frac{n}{n-1}} MemoryOperations}{(3M)^{\frac{n}{n-1}}} - 1)
\end{equation*}

\begin{equation*}
    TotalRead = 2(\frac{n}{3})^{\frac{n}{n-1}}(\frac{MemoryOperations}{M^{\frac{1}{n-1}}} - M)
\end{equation*}

\subsection{GEMV I/O Lower Bound}

The total number of memory operations is $mn$ if the $m$ and $n$ are the problem size of the operation. 

There are three Linear Algebra objects which read from the slow memory to the fast memory. Which gives

\begin{align*}
    TotalRead &= 2(\frac{3}{3})^{\frac{3}{3-1}}(\frac{mn}{M^{\frac{1}{3-1}}} - M)\\
    TotalRead &= (\frac{2mn}{\sqrt{M}} - 2M)
\end{align*}

To get the least amount of read operations on the machine which supports fast of size $M$, we have to consider $m$ initial read operation for vector, $c$.

\begin{equation}
    \frac{2mn}{\sqrt{M}} - m - 2M
    \label{eq:gemv_least_read}
\end{equation}

To get the least amount of store operations on the machine which supports fast of size $M$, we have at least $m - M$ compulsory stores for vector, $c$.

\begin{equation}
    \frac{2mn}{\sqrt{M}} + m - 3M
    \label{eq:gemv_least_store}
\end{equation}

\subsection{GEM I/O Lower Bound}

The total number of memory operations is $mn$ if the $m$ and $n$ are the problem size of the operation. 

There are three Linear Algebra objects which read from the slow memory to the fast memory. Which gives

\begin{align*}
    TotalRead &= 2(\frac{3}{3})^{\frac{3}{3-1}}(\frac{mn}{M^{\frac{1}{3-1}}} - M)\\
    TotalRead &= (\frac{2mn}{\sqrt{M}} - 2M)
\end{align*}

To get the least amount of read operations on the machine which supports fast of size $M$, we have to consider $mn$ initial read operation for matrix, $C$.

\begin{equation}
    \frac{2mn}{\sqrt{M}} - mn - 2M
    \label{eq:gem_least_read}
\end{equation}

To get the least amount of store operations on the machine which supports fast of size $M$, we have at least $mn - M$ compulsory stores for matrix, $C$.

\begin{equation}
    \frac{2mn}{\sqrt{M}} + mn - 3M
    \label{eq:gem_least_store}
\end{equation}

\subsection{DOT I/O Lower Bound}

The total number of memory operations is $n$ if the $n$ is the problem size of the operation. 

There are two Linear Algebra objects and one scalar which read from the slow memory to the fast memory. Which gives

\begin{align*}
    TotalRead &= 2(\frac{2}{3})^{\frac{2}{2-1}}(\frac{mn}{M^{\frac{1}{2-1}}} - M)\\
    TotalRead &= \frac{1}{3}(\frac{8n}{M} - 8M)
\end{align*}

To get the least amount of read operations on the machine which supports fast of size $M$, we have to consider $1$ initial read operation for scalar, $\alpha$.

\begin{equation*}
    \frac{1}{3}(\frac{8n}{M} - 8M) - 1
\end{equation*}

\begin{equation}
    \frac{1}{3}(\frac{8n}{M} - 8M)
    \label{eq:dot_least_read}
\end{equation}

To get the least amount of store operations on the machine which supports fast of size $M$, we have at least $1 - M$ compulsory stores for scalar, $\alpha$.

\begin{equation*}
    \frac{1}{3}(\frac{8n}{M} - 8M) + 1 - M
\end{equation*}

\begin{equation}
    \frac{8n}{3M} - \frac{11M}{3}
    \label{eq:dot_least_store}
\end{equation}

\section{Algorithm and Analysis}

This section will first give the algorithm, and then we will analyse and compare the I/O lower bound.

\subsection{Column-Major GEMV Algorithm}

The GEMV algorithm based on the algorithm defined for the Matrix-Matrix product by BLIS and all the tunning parameters calculated using the methods defined in \cite{BLIS}. This algorithm requires only two levels of cache, which makes the calculation of the parameters much more straightforward.

\begin{algorithm}[H]
    \SetAlgoLined

    \Begin{
        \tcp{Loop 1}
        \For{\assign{i_c}{0} \KwTo $m$ \KwBy $m_c$}{
            \tcp{Loop 2}
            \For{\assign{j_c}{0} \KwTo $n$ \KwBy $n_c$}{
                \tcp{Loop 3}
                \For{\assign{i_r}{0} \KwTo $m_c$ \KwBy $m_r$}{
                    \tcp{Micro-Kernel}
                    \assign{c(i_r : i_r + m_r)}{c(i_r : i_r + m_r) + A(i_r : i_r + m_r, j_c: j_c + n_c) \times x(j_c: j_c + n_c)}
                }   
            }
        }
    }

    \caption{Column-Major Matrix-Vector Product}
\end{algorithm}

The block $A$ of size $m_c \times n_c$ resides inside the $L_2$ cache, the block of $b$ resides inside the $L_1$, and the block $c$ is brought from the slow memory into registers.

Three opportunities are available for parallelization: Loop 1, Loop 2, and Loop 3. Here, we have to consider Amdahl's Law during parallelizing loop because its effect starts to visible for enormous problem set size, and choose a large percentage of code to be parallelized.

Let $p$ be the percentage of parallelizable code, $p - 1$ is a part that is not parallelizable, $T$ is time spent inside the serial code, $t$ be the number of threads, and $O(t)$ be the thread overhead.

\begin{align*}
  T_{total} &= pT + (1-p)T\\
  T_{total_t} &= \frac{pT}{t} + (1-p)T\\
\end{align*}

Assuming super-linear speedup is not expected then
\begin{equation}
    T_{total_t} \ge \frac{pT}{t} + (1-p)T \ge (1-p)T\\
\end{equation}

\begin{equation*}
    S_t = \frac{T_{total}}{T_{total_t}} - O(t)
\end{equation*}

\begin{equation}
    S_t \le \frac{1}{1-p} - O(t)
    \label{eq:speedup}
\end{equation}

Therefore, to maximize $S_t$, we have to maximize parallelize large amount of serial code. This is true for for all Level-1 and Level-2 BLAS operations.

If we parallelize a loop, we have to consider which part we spend most of our time on. The time spent goes like Loop-1 $>$ Loop-2 $>$ Loop3 because Loop-1 has a larger block size and spends more time during computation. However, as the size of the problem increase, we start to see the effect of bandwidth.

We decoupled the GEMV algorithm based on the layout type because of the way this operation is defined. For example, to get the element for each row in the vector, $c$, we need to take a dot product with each row vector from matrix $A$ to the whole vector, $b$. Therefore,  the column-major GEMV has a very bad cache locality because the stride is not one, whereas the row-major GEMV has an excellent locality due to the stride being one, which the CPU loves to have.

\subsubsection{Analysis}

The number of memory access incurred by the algorithm for each level of cache:

\textit{Blocking for $L_2$ cache}. In our algorithm, the dimension with size m is partition into the block of size $m_c$, and n is partition into the block of size $n_c$ and determined by $L_1$ cache. This allows us to put the block of matrix $A$ in the $L_2$ cache. The block of vector, $x$ and the vector, $c$, also need to use the $L_2$ cache to get inside the $L_1$ cache and registers.
\[\frac{mn}{m_c} + \frac{mn}{n_c} + m\]
The solution to the tunning parameter is close to the optimal because the $m_c \approx n_c$, which points us towards the matrix A being a square matrix. Since, $m_c \approx n_c \approx \sqrt{M_{L_2}}$, we get
\[\frac{mn}{m_c} + \frac{mn}{n_c} \approx \frac{mn}{\sqrt{M_{L_2}}} + \frac{mn}{\sqrt{M_{L_2}}}\]
\[\frac{mn}{m_c} + \frac{mn}{n_c} \approx \frac{2mn}{\sqrt{M_{L_2}}}\]

\textit{Blocking for $L_1$ cache}. Here, partition in $m$ dimension is $m_r$, which is much smaller than the $\sqrt{M_{L_1}}$ or $.m_r \ll \sqrt{M_{L_1}} $. This makes it sub-optimal and we don't partition $n$ dimension, $n_c > \sqrt{M_{L_1}} \gg m_r$

\[\frac{mn}{m_r} + \frac{mn}{n_c} \approx \frac{mn}{m_r}\]
\[\frac{mn}{m_r} > \frac{2mn}{\sqrt{M_{L_1}}}\]

\subsection{Row-Major GEMV and GER Algorithm}

\begin{algorithm}[H]
    \SetAlgoLined

    \Begin{
        \tcp{Loop 1}
        \For{\assign{i}{0} \KwTo $n$ \KwBy $1$}{
            \tcp{Loop 2}
            \For{\assign{j}{0} \KwTo $m$ \KwBy $1$}{
                \tcp{Operation based on the type of algorithm}
                \tcp{For GEMV: $c_i \gets c_i + A_{ij} \times b_j$}
                \tcp{For GER: $C_{ij} \gets C_{ij} + a_i \times b_j$}
            }
        }
    }
    \caption{GER or Row-Major GEMV}
\end{algorithm}

\subsubsection{Analysis}

Both algorithms have an excellent cache locality, which makes the blocking inefficient compared to their without blocking counterpart. Therefore, most of the libraries miss the opportunity to optimize these operations. For both, the memory operation is sub-optimal compared to our I/O lower bound, but the memory being linear gives the CPU to use cache locality and fetch a whole cache line. The fetching of the cache line increases the hit ratio. There is a high probability that it is present inside the cache line when asking for the next element. The CPUs have been mastering this behaviour for decades and optimizing the memory operation for linear data.

There is only one type of optimization that most libraries are missing, and we did get a great result. We got the performance boost compared to Intel's MKL, even though we used the OpenMP to vectorize code.

We can use a blocking algorithm due to the linear data; therefore, we propose to add multi-threading support to the algorithm. Here, we have two opportunities to add multi-threading, but we have to consider the thread overhead that is quite noticeable for small before using the threads. We get worse performance for small problem size because the parallelizable part is small and thread overhead is high. The thread overhead, $O(t)$ dominates the speedup we get from parallelizing the code and can be seen in the equation \ref{eq:speedup}. Although the thread overhead dominates smaller problem size operations, and it is also a function of threads—the overhead increases as the number of threads increases. 

There are two possible solutions to this problem:
\begin{enumerate}
    \item We take a small number of threads for smaller problem.
    \item We do not enable multi-threading for smaller problem.
\end{enumerate}

In our implementation to increase the data locality, we iterate over the column in Loop-1, and this allows us to iterate in a sequential linearly laid memory vectors and matrix. Furthermore, we hand over each column to each thread, which is only possible when the Loop-1 parallelized. On the other hand, if we parallelize Loop-2, there is a data race in GER, and also it increases the cache miss ratio. 

\subsection{Dot Algorithm}

\begin{algorithm}[H]
    \SetAlgoLined

    \Begin{
        \tcp{Micro-Kernel}
        \assign{\alpha}{\alpha + a(0:m) \times b(0:m)}
    }
    \caption{Small Problem Size Dot Product}
\end{algorithm}

\begin{algorithm}[H]
    \SetAlgoLined

    \Begin{
        \tcp{Loop-1}
        \For{\assign{i}{0} \KwTo $n$ \KwBy $n_c$}{
            \tcp{Micro-Kernel}
            \assign{\alpha}{\alpha + a(i: i + m_c) \times b(i: i + m_c)}
        }
    }
    \caption{Large Problem Size Dot Product}
\end{algorithm}

Here, again there is an opportunity to utilize modern hardware, and most of the libraries ignored it because of its simplicity. The data is also laid sequentially in memory that allows the CPU to give the best possible performance without blocking. The problem with the current implementation with the popular libraries is that they do not utilize the multi-core features of the CPU except \textbf{Intel's MKL}. Now, we know we need to apply multi-threading, but how? There is no exposed loop that is parallelizable with the current implementation. That is why we explicit loop and wrap it around our micro-kernel with a block size $n_c$.

\subsubsection{Analysis}

We have two different implementations based on the size of the problem because, for a smaller problem size, the thread overhead completely dominates the performance, which is possible with the single-core. We get worse performance for a smaller size because all the data can stay inside the $L_1$ cache, and the hit ratio is pretty high. There are almost no misses, but when we introduce the threads, we have to pay the cost of creating threads, and the data that can fit inside the $L_1$ cache of a single-core now needs to be divided among different cores, remember these $L_1$ caches are not shared; therefore, all the cores start to ask for data and increase the misses. The best cutoff point for using the smaller size dot product and the larger size product is when the problem size is greater than the $L_1$ cache. If the problem size is less than the $L_1$ cache, use the smaller size dot product; otherwise, the larger size dot product. The lower bound for the cutoff was found using experiments and also be derived using simple logic. When the data size is smaller, the threads will contribute to overhead until the data can no longer fit inside the cache, and when the data size is more significant than cache size, the misses will increase and overtake thread overhead.

\[DotProductCutoff \ge M_{L_1}\]

In our implementation, dot product cutoff point that we choose is $2M_{L_1}$.

\textit{Block Size for $L_1$ cache}. Our bottle-neck is not the flops but memory operations because each core needs the data to be present in its cache, and flops can be optimized inside the micro-kernel, and it is not our focus. Our primary focus is to parallelize the dot product; therefore, there are $2n$ memory operations, and to get the block size, we know that at any point in time, $2n_c$ number of elements needs to be present inside the $L_1$ cache for each core. So we will this fact and derive the block size for the larger size dot product.

\[S_{data}(2n_c) = S_{L_1}\]
\[n_c = \frac{S_{L_1}}{2S_{data}}\]
\[n_c = \frac{M_{L_1}}{2}\]

Now, let us check the I/O lower bound on the larger size dot product.

\[\frac{n}{n_c} + 1 \approx \frac{2n}{n_c}\]
\[\frac{n}{\frac{M}{2}}\]
\[\frac{2n}{M} \approx \frac{8n}{3M}\]
\[\frac{2n}{M} \approx \frac{2.67n}{M}\]

Therefore, our algorithm is optimal.
\section{Performance}

\FloatBarrier
\begin{figure}[ht!]
    \centering
    \caption{Small Size Dot Product Performance Plot}
    \includegraphics[width=12cm]{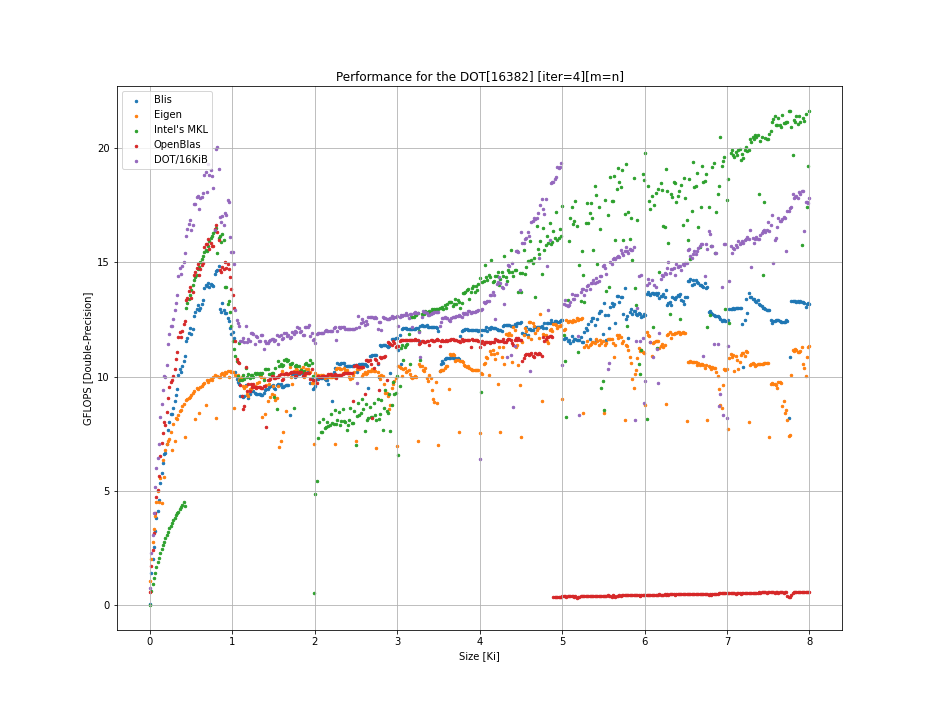}
\end{figure}
\begin{figure}[ht!]
    \centering
    \caption{Larger Size Dot Product Performance Plot}
    \includegraphics[width=12cm]{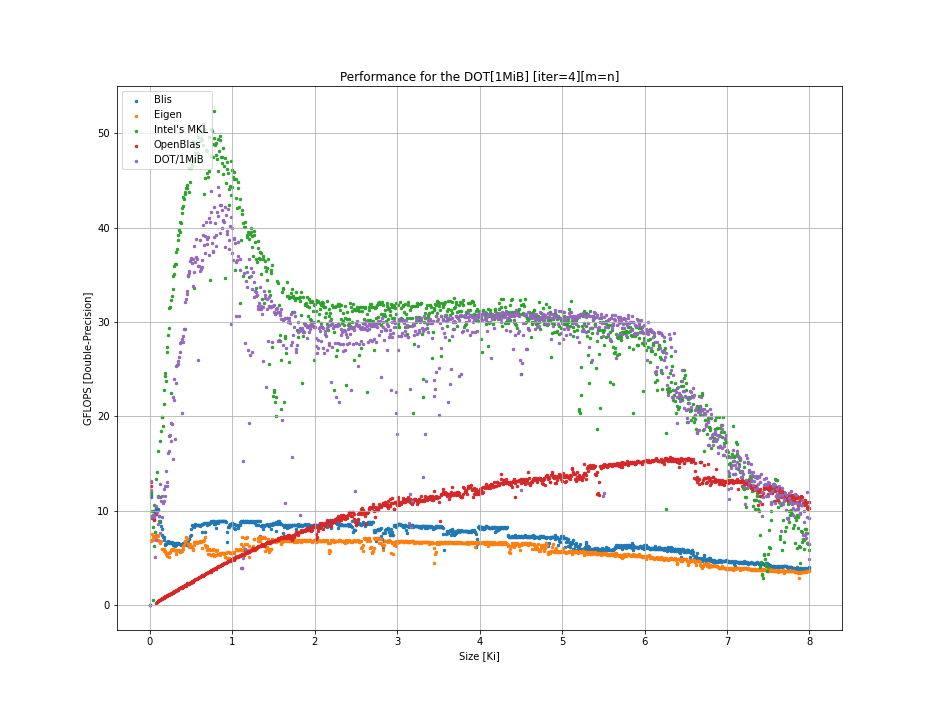}
\end{figure}
\begin{figure}[ht!]
    \centering
    \caption{GER Performance Plot}
    \includegraphics[width=12cm]{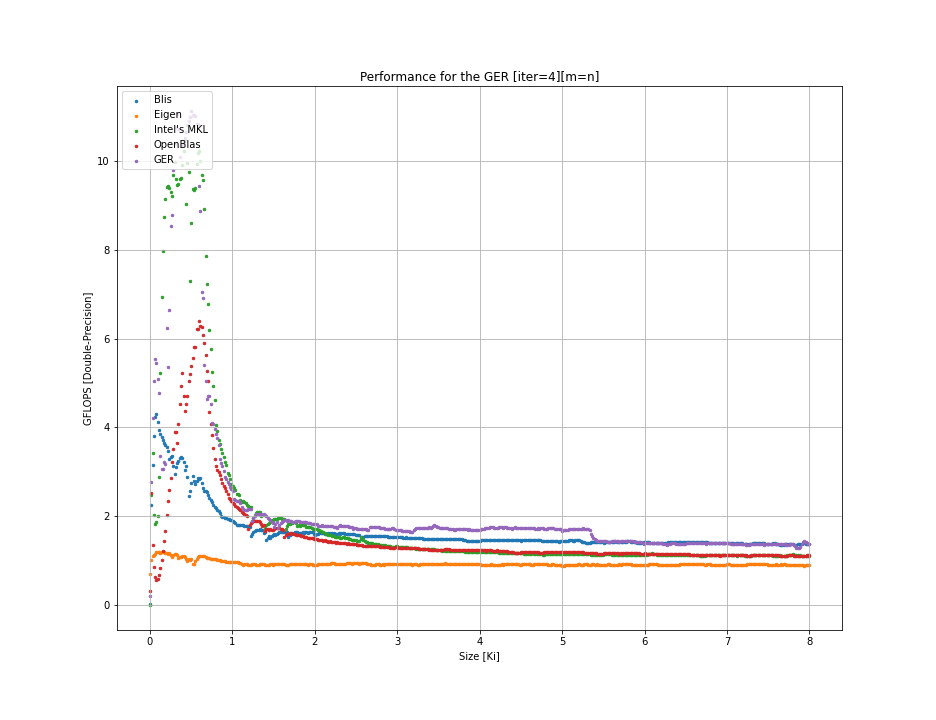}
\end{figure}
\begin{figure}[ht!]
    \centering
    \caption{Row-Major GEMV Performance Plot}
    \includegraphics[width=12cm]{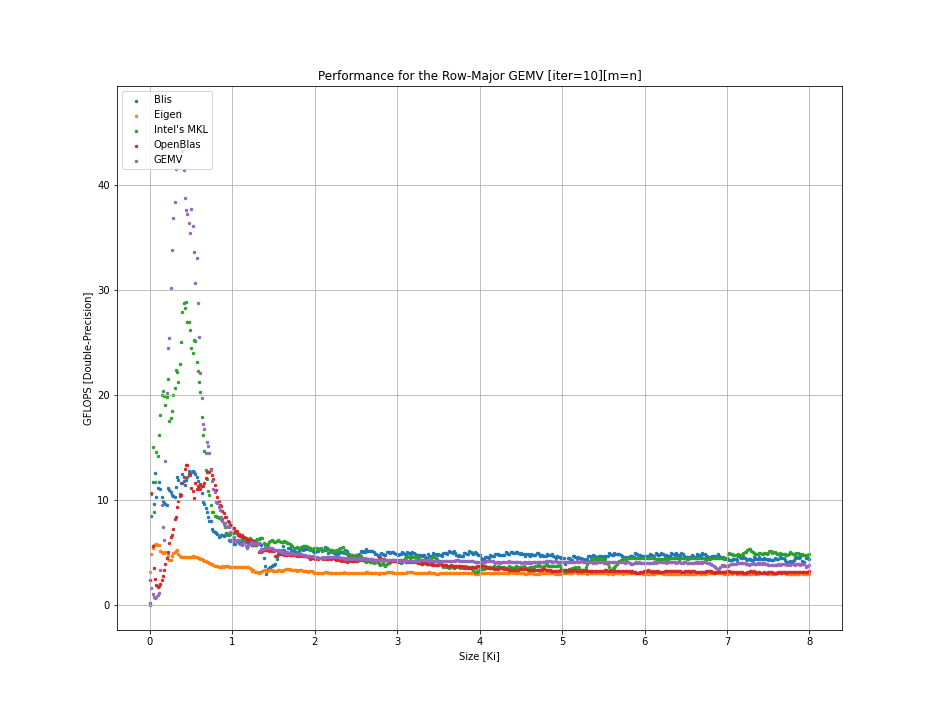}
\end{figure}
\begin{figure}[ht!]
    \centering
    \caption{Column-Major GEMV Performance Plot}
    \includegraphics[width=12cm]{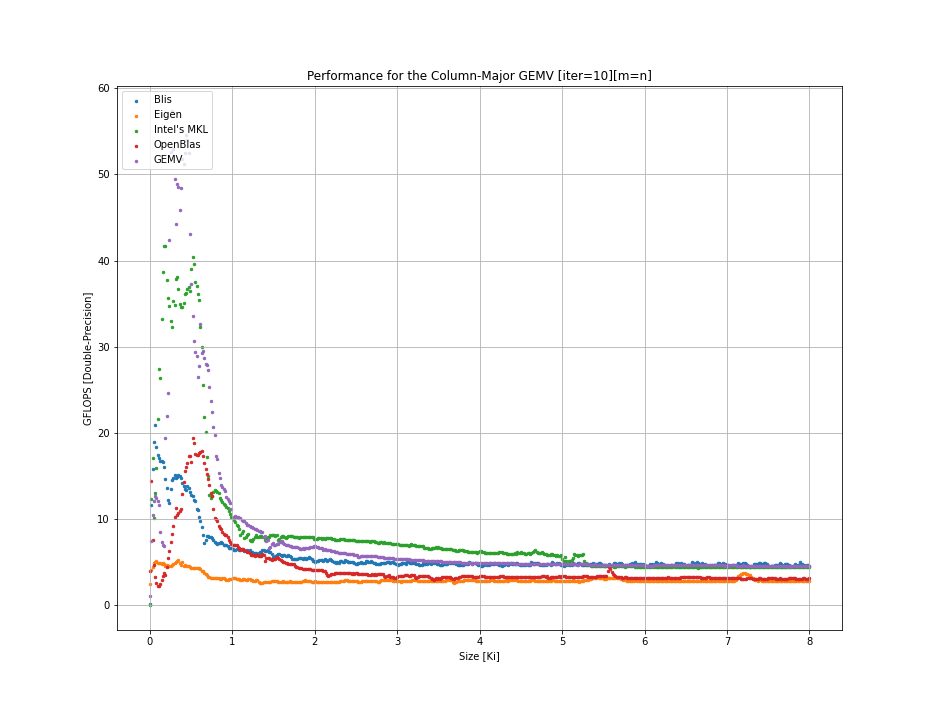}
\end{figure}
\FloatBarrier
\newpage

\bibliographystyle{unsrtnat}
\bibliography{references}
\end{document}